\documentclass[letterpaper]{jpconf}
\usepackage{graphicx,epsfig}
\def\be{\begin{equation}}
\def\ee{\end{equation}}
\def\bea{\begin{eqnarray}}
\def\eea{\end{eqnarray}}
\def\pp{\psi(2S)}
\def\jp{J/\psi}
\begin{document}
\title{Hadronic Dynamics in the Studies at B Factories}

\author{M.B. Voloshin}

\address{William I. Fine Theoretical Physics Institute, University of
Minnesota, Minneapolis, MN 55455, USA \\
and \\
Institute of Theoretical and Experimental Physics, Moscow, 117218, Russia}

\begin{abstract}
I discuss topics in the strong QCD dynamics that are of relevance to studies of the $B$ mesons and to the spectroscopy of heavy hadrons. In particular the limitations from the hadronic dynamics for the determination of the weak-interaction parameters are discussed as well as some models for the newly observed apparently exotic resonances with hidden charm.
\end{abstract}

\section{Introduction}
The studies of the parameters of the Standard Model and the searches for a `New Physics' in the properties of $B$ mesons inevitably run into the usual problem that we formulate our theories in terms of quarks and gluons, while the experimental data are collected in terms of hadrons. Thus an interpretation of the wealth of data accumulated at the B factories within the underlying theory crucially depends on our ability to cope with the strong dynamics and to `translate' between the quark-gluon and hadronic `languages'. In this lecture I discuss some known results in such a `translation' and the existing uncertainties, which still limit the precision of extracting the theoretical parameters. Some of these uncertainties can be somewhat relaxed with a help from new data, however some still appear to be beyond the current understanding of the strong dynamics. In what follows I discuss the current theoretical uncertainties in the precision of determination of the weak-interaction parameters from the data and, in some cases, the ways of reducing such uncertainties by further phenomenological studies of the yet unknown theoretical parameters resulting from the strong dynamics. 

The studies at the B factories thus far have not produced any credible indications of a `New Physics'. However the unprecedented amount of high precision data accumulated in the process may be a treasure trove for learning new aspects of the QCD dynamics of quarks and leptons. In particular, one unexpected spin-off from the experiments at the B factories is the observation of a multitude of heavy resonances, both with open and hidden charm, displaying highly unusual properties, apparently indicating that some of those are the much expected exotic states, i.e. their structure goes beyond the standard quark-antiquark template for mesons. At present it looks like the new resonances fall into different categories according to their different internal structure. I discuss here two such types: the apparently `molecular' state $X(3872)$ and a class of resonances ($Y$, $Z$) which decay into specific charmonium states and pions. I describe in some detail the model, where these latter states are that of a `hadro-charmonium', i.e. they can be considered as a compact charmonium state embedded in excited light-matter.

\section{Weak interaction parameters}
The weak decays of $b$ hadrons provide us with the access to measurements of the entries in the Cabibbo - Kobayashi - Maskawa mixing matrix, describing the weak interaction mixing of the third quark generation.
\subsection{ $|V_{cb}|$}
\subsubsection{$V_{cb}$ from inclusive decay rate of $B \to X_c \ell \nu$.}
The most straightforward technique for determining the mixing parameter $|V_{cb}|$ is to apply the essentially parton picture to the rate of inclusive semileptonic decay of a $b$ hadron, i.e. to identify the decay rate of a hadron, e.g. $B \to X_c \ell \nu$, with the quark decay rate due to the process $b \to c \ell \nu$, modulo small nonperturbative terms that take into account the `Fermi motion' of the $b$ quark  and also the $b$ quark spin interaction with the chromomagnetic field that the heavy quark `sees' inside the hadron. Those nonperturbative terms are suppressed by the parameter $\Lambda_{QCD}^2/m_b^2$ and can be treated\cite{bigi92} within an application of the Operator Product Expansion to the inclusive decay rates\cite{shifman84,shifman86}. Certainly, such approach assumes that the $b$ quark mass is large enough for the uncontrollable low-energy non-perturbative effects to die out and thus for the quark-hadron duality to set in. It is not known at present to which accuracy this assumption works, however to some extent it is tested in the treatment of the lifetimes of $b$ hadrons and their differences, as will be discussed further in this lecture. One trivial cross-check of such approach is to verify its prediction that in absolute terms the inclusive semileptonic decay rates $B \to X_c \ell \nu$ should be the same. The data agree with this expectation very well: according to the Tables \cite{pdg}, one can deduce  
$\Gamma(B^+ \to \ell^+ \nu X) = 67.1 \pm 1.8 \, ns^{-1}$~~~~$\Gamma(B^0 \to \ell^+ \nu X)  = 67.5 \pm 1.8 \, ns^{-1}$. The charmless decay $b \to u \ell \nu$ contributes only about 2\% to each of these rates, so that this small contribution can be treated in either way without introducing an uncertainty exceeding the experimental error.
The theoretical expression for the decay rate thus has the generic form (clearly related to the familiar muon decay expression)
\be
\Gamma(B \to \ell \nu X_c) = {G_F^2 { |V_{cb}|^2} \, m_b^5 \over 192 \pi^3} \, f \left ({m_c \over m_b}\right ) \, \times \left [ 1 + (pert.~ correct) + (non-pert. ~correct) \right ]~,
\label{gsli}
\ee
with the well known kinematical function
\be
f(x)=1-8x^2 + 8 x^6 - x^8 - 24x^4 \, \ln x~.
\ee
The explicit form of the corrections depends on the scheme used in a definition of the quark masses $m_b$ and $m_c$ \cite{bigi97,hoang98,uraltsev04,aquila05}. The latter mass parameters and the uncertainty in them in fact dominate the problem of extracting $|V_{cb}|$ from the data. One approach\cite{mv94} to this problem is to fit these kinematical quantities from the data on the spectra in the semileptonic $B$ decays, e.g. from the moments of the charged lepton energy $E_\ell$: $M_n = \int E_\ell^n \, {d \Gamma_{sl} \over d E_\ell} \, d E_\ell$ (in this classification $\Gamma_{sl}=M_0$). The difficulty of this approach is in the nonuniform over the spectrum experimental sensitivity and in that higher moments are generally less certain theoretically due to larger corrections and also due to enhanced potential importance of the unknown deviations from the quark-hadron duality. A typical fit to the available data along these lines yields (see e.g. in \cite{hfag}) $|V_{cb}| \approx (41.7 \pm 0.8) \times 10^{-3}$, where the error includes the experimental errors and the error of the fit but does not include any theoretical uncertainty.  

\subsubsection{$|V_{cb}|$ from exclusive decays $B \to D^{(*)} \ell \nu$ at zero recoil.}
An alternative way of extracting the mixing parameter $|V_{cb}|$ is provided by the prediction from the heavy quark symmetry for the form factors of the decays $B \to D^* \ell \nu$ and $B \to D \ell \nu$ at zero recoil\cite{shifman87}, i.e. when the final charmed meson is produced at rest. Namely at the kinematical point where $w \equiv (p_B \cdot p_D)/ (M_B M_D) =1$ the axial form factor $F$:
\be
\langle D^*(\epsilon)| ({\bar c} \gamma_5 \gamma_\mu b) | B \rangle = \epsilon_\mu \, F(1)
\label{aform}
\ee
and the vector form factor $G$:
\be
\langle D| ({\bar c}  \gamma_0 b) | B \rangle = G(1)
\label{vform}
\ee
are both close to one with only perturbative correction and corrections to the heavy quark limit:
$F(1), G(1) = 1 +$ pert.corr. $+O(m_c^{-2}, m_b^{-2})$. The averages\cite{hfag} of the experimental results for the measurable products of these form factors with $|V_{cb}|$ are $F(1) \, |V_{cb}|= (35.97 \pm 0.53) \times 10^{-3}$ and $G(1) \, |V_{cb}|= (42.3 \pm 4.5) \times 10^{-3}$. Thus the real issue is the theoretical estimate of how far the form factors deviate from one as a result of the corrections. The axial form factor $F(1)$ is estimated to be about 0.92 both analytically\cite{bigi97} and numerically (on a lattice)~\cite{laiho07} with an error believed to be of few percent, where `few' typically implies 3 to 6. This estimate results in $|V_{cb}| = [{ 39.1 \pm 0.6_{\rm exp} \pm (1.1 \div 2.2)_{\rm th}}] \times 10^{-3}$.
The vector form factor is estimated numerically\cite{okamoto04} as $G(1) \approx 1.07$, which results in $|V_{cb}| = [39.1 \pm 4.2_{\rm exp} \pm 0.9_{\rm lat}] \times 10^3 $.

Although the two exclusive determinations of $|V_{cb}|$ are in a perfect agreement with one another and also agree, within the uncertainties, with the inclusive determination, the estimates for the form factors may raise some reservations. The reason is that additionally to the heavy quark symmetry the closeness of $G(1)$ (but not of $F(1)$) to one is also protected by the Ademollo-Gatto theorem\cite{ag64} in the limit $m_c \to m_b$. In the real world the masses of the charm and bottom quarks are of course different, however the relevant parameter breaking the symmetry limit in $G(1)$ is $\xi^2 \equiv [(m_b-m_c)/(m_b+m_c)]^2 \approx 0.25$~\cite{shifman87}, and one could expect that some `remnants' of the symmetry suppress the difference $|G(1)-1|$ in comparison with the `unprotected' $|1-F(1)|$. Clearly, the currently used estimates of the deviations of the form factors from one show no traces of such a suppression. It can be noticed that if the actual values of the form factors indeed satisfy the Ademollo-Gatto requirement literally: $G(1)-1 \approx \xi^2 \, [1 - F(1)]$, all three determinations of $|V_{cb}|$ would naturally `click' into place, noticeably better than they are currently believed to. Namely, with $\xi^2=0.25$ one finds
\be
|V_{cb}| = {4 \, G(1) \, |V_{cb}| + F(1) \, |V_{cb}| \over 5} = 41.0 \pm 3.6
\label{vcb}
\ee
in a perfect agreement with the determination from the inclusive semileptonic decay. The data then also imply that $F(1) \approx 0.88$ and $G(1) \approx 1.03$.

It can be also mentioned that an agreement between all three determinations of $|V_{cb}|$ can be viewed as a test of the $V-A$ structure of the $b \to c$ weak current. In particular the limit on the ratio of the right- and left-chiral amplitudes $|V+A|/|V-A|$ is about 0.15~\cite{mv97}.

\subsection{The semileptonic branching fraction ${\cal B}_{sl}(B)$.}
Although it is not directly related to the determination of the CKM parameters, the low measured overall semileptonic branching fraction ${\cal B}_{sl}(B)$ has been challenging to the theoretical understanding of the weak decays of the $B$ mesons\cite{bigi93}. The nonperturbative effects (in the hadronic decay) are smaller for the $B^0$ meson. Thus the experimental number appropriate for a use as a reference is ${\cal B} (B^0 \to \ell^+ \nu X)= (10.33 \pm 0.28)\%$. On the theoretical side ${\cal B}_{sl} (B^0)$ is determined dominantly by the `parton' decay branching ratio:
\be
{\cal B}_{sl}(b \to c e \nu)=
{\Gamma(b \to c e \nu) \over \sum_{\ell = e, \mu, \tau} \Gamma(b \to c \ell \nu) + \Gamma(b \to c {\bar u} d(s)) + \Gamma (b \to c {\bar c} s (d))+\Gamma(b \to u \, X) + \Gamma_{\rm rare}}
\label{bsl}
\ee
with the main factor determining the ratio being the ratio of the quark and the semileptonic decay rates of the $b$ quark:
\be
{\Gamma(b \to c {\bar u} d(s)) \over 3 \, \Gamma(b \to c e \nu)} = 1 + {\alpha_s \over 2 \pi}+ {\alpha_s^2 \over \pi^2} \, 4 \, \left (L^2 + {15 \over 8} \, L + { 1.3} + {1.8} \right )+ \ldots~,
\label{bslt}
\ee
where $L=\ln(m_W/m_b) \approx 2.8$. The calculation\cite{acz06} of NNLO terms in the $O(\alpha_s^2)$ radiative correction has increased the predicted relative hadronic decay rate and thereby lowered the prediction for the overall semileptonic branching fraction from 11.5\% to 11\%. As a result, the remaining gap between the theory and the data does not look as baffling\cite{bigi93} as it used to.

\subsection{Lifetime differences.}
\begin{figure}[ht]
\unitlength 0.7mm
\thicklines
\begin{picture}(140.00,84.00)(-20,0)
\put(10.00,70.00){\line(1,0){20.00}}
\put(30.00,70.00){\line(1,0){30.00}}
\put(60.00,70.00){\line(1,0){20.00}}
\bezier{164}(30.00,70.00)(45.00,84.00)(60.00,70.00)
\bezier{164}(30.00,70.00)(45.00,56.00)(60.00,70.00)
\put(30.00,70.00){\circle*{2.00}}
\put(60.00,70.00){\circle*{2.00}}
\put(20.00,72.00){\makebox(0,0)[cb]{$Q$}}
\put(70.00,71.00){\makebox(0,0)[cb]{$Q$}}
\put(49.00,78.00){\makebox(0,0)[cb]{$q_1$}}
\put(49.00,71.00){\makebox(0,0)[cb]{$q_2$}}
\put(49.00,62.00){\makebox(0,0)[ct]{${\overline q_3}$}}
\put(10.00,43.00){\line(1,0){20.00}}
\put(30.00,43.00){\line(1,0){30.00}}
\put(60.00,43.00){\line(1,0){20.00}}
\bezier{164}(30.00,43.00)(45.00,57.00)(60.00,43.00)
\bezier{164}(30.00,43.00)(45.00,29.00)(60.00,43.00)
\put(30.00,43.00){\circle*{2.00}}
\put(60.00,43.00){\circle*{2.00}}
\put(20.00,45.00){\makebox(0,0)[cb]{$Q$}}
\put(70.00,44.00){\makebox(0,0)[cb]{$Q$}}
\put(49.00,51.00){\makebox(0,0)[cb]{$q_1$}}
\put(49.00,44.00){\makebox(0,0)[cb]{$q_2$}}
\put(49.00,35.00){\makebox(0,0)[ct]{${\overline q_3}$}}
\put(45.00,36.00){\line(0,-1){2.00}}
\put(45.00,33.00){\line(0,-1){2.00}}
\put(45.00,30.00){\line(0,-1){2.00}}
\put(46.00,28.00){\makebox(0,0)[cc]{$g$}}
\put(10.00,12.00){\line(1,0){20.00}}
\put(30.00,12.00){\line(1,0){30.00}}
\put(60.00,12.00){\line(1,0){20.00}}
\bezier{164}(30.00,12.00)(45.00,26.00)(60.00,12.00)
\put(30.00,12.00){\circle*{2.00}}
\put(60.00,12.00){\circle*{2.00}}
\put(20.00,14.00){\makebox(0,0)[cb]{$Q$}}
\put(70.00,13.00){\makebox(0,0)[cb]{$Q$}}
\put(49.00,20.00){\makebox(0,0)[cb]{$q_1$}}
\put(49.00,13.00){\makebox(0,0)[cb]{$q_2$}}
\put(10.00,6.00){\line(3,1){20.00}}
\put(60.00,12.00){\line(4,-1){20.00}}
\put(20.00,7.00){\makebox(0,0)[ct]{$q_3$}}
\put(70.00,8.00){\makebox(0,0)[ct]{$q_3$}}
\put(87.00,70.00){\vector(1,0){11.00}}
\put(87.00,43.00){\vector(1,0){11.00}}
\put(87.00,12.00){\vector(1,0){11.00}}
\put(110.00,70.00){\line(1,0){30.00}}
\put(116.00,71.00){\makebox(0,0)[cb]{$Q$}}
\put(134.00,71.00){\makebox(0,0)[cb]{$Q$}}
\put(110.00,43.00){\line(1,0){30.00}}
\put(116.00,44.00){\makebox(0,0)[cb]{$Q$}}
\put(134.00,44.00){\makebox(0,0)[cb]{$Q$}}
\put(125.00,43.00){\line(0,-1){2.00}}
\put(125.00,40.00){\line(0,-1){2.00}}
\put(125.00,37.00){\line(0,-1){2.00}}
\put(126.00,35.00){\makebox(0,0)[cc]{$g$}}
\put(110.00,12.00){\line(1,0){30.00}}
\put(116.00,13.00){\makebox(0,0)[cb]{$Q$}}
\put(134.00,13.00){\makebox(0,0)[cb]{$Q$}}
\put(110.00,7.00){\line(3,1){15.00}}
\put(125.00,12.00){\line(3,-1){15.00}}
\put(116.00,8.00){\makebox(0,0)[lt]{$q$}}
\put(132.00,9.00){\makebox(0,0)[rt]{$q$}}
\put(125.00,62.00){\makebox(0,0)[cc]{ $m_Q^5 \, ({\overline Q} Q)$}}
\put(125.00,27.00){\makebox(0,0)[cc]{ $ m_Q^3 \, \left ({\overline Q}
(\vec \sigma \cdot \vec B) Q \right )$}}
\put(125.00,0.00){\makebox(0,0)[cc]{ $m_Q^2 \, ({\overline q} \Gamma
q)({\overline Q} \Gamma^\prime Q)$}}
\put(125.00,70.00){ \circle*{2.83}}
\put(125.00,43.00){ \circle*{2.83}}
\put(125.00,12.00){ \circle*{2.83}}
\end{picture}
\caption{The three types of graphs whose unitary cuts generate the first three terms of the heavy quark expansion for the inclusive weak decay rates.}
\label{figope}
\end{figure}
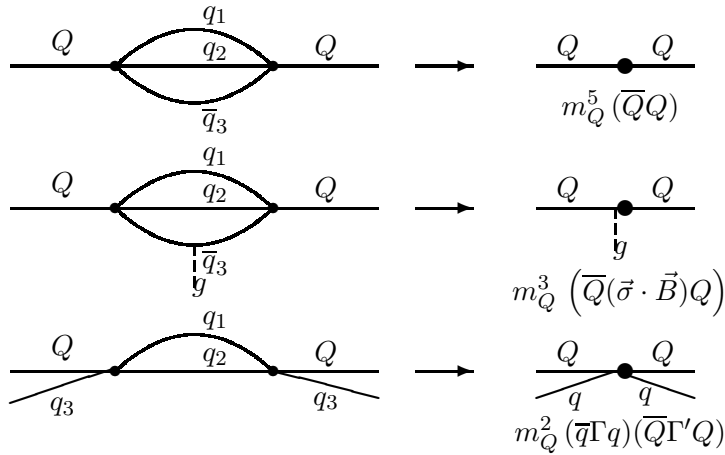

The study of of the differences between inclusive decay rates of different $b$ hadrons provides an insight into the applicability of using the heavy quark limit and the operator product expansion (OPE) in the Minkowski domain, i.e. the validity of the presumed quark-hadron duality. Within the OPE based approach the inclusive decay rate is expressed by means of the effective Lagrangian, which in fact is the absorptive part of the correlator of two (appropriate parts of the) weak interaction Lagrangians:
\be
L_{eff}=2 \,{\rm Im} \, \left [ i \int d^4x \, e^{iqx} \, T \left \{
L_W(x),
L_W(0) \right \} \right ]~,
\label{leff}
\ee
so that the decay rate of a heavy hadron $H_Q$ is calculated as
\be
\Gamma_H=\langle H_Q | \, L_{eff} \, | H_Q \rangle~,
\label{gleff}
\ee
at $q^2=m_Q^2$.
For a heavy quark one can use the OPE in the inverse powers of the heavy quark mass $m_Q$~\cite{shifman84,shifman86}:
\bea
&&L_{eff}= L_{eff}^{(0)} + L_{eff}^{(2)} + L_{eff}^{(3)}=\nonumber \\
&&{ const \cdot {G_F^2 \, m_Q^5 \over 192 \, \pi^3} \, \left ( {\overline Q} Q \right )} + { c_2 \, m_Q^3 \, \left ({\overline Q}
(\vec \sigma \cdot \vec B) Q \right )} + { \sum_i c_3^{(i)} \, m_Q^2 \,
({\overline q}_i \Gamma_i
q_i)({\overline Q} \Gamma^\prime_i Q)}+\ldots~,
\label{opem}
\eea
where the first term describes the perturbative `parton' decay, the second term introduces a correction due to the local gluonic field inside the hadron and the third term describes the effect of the spectator quark in the decay rate. The three first terms of the expansion in $m_Q^{-1}$ correspond to the unitary cuts of the diagrams shown in Figure~\ref{figope}.

The difference in the inclusive decay rates of the $b$ hadrons with different light-quark flavor arises from the third term of the expansion (\ref{opem}). In particular one finds\cite{shifman86,run2} that the lifetimes of the $B^0_d$ and $B_s$ should be the same within less than one percent, while the lifetime of the $B^+$ meson should be measurably longer:
\be
\Gamma(B^\pm) - \Gamma(B^0) = |V_{cb}|^2 \,{G_F^2 \, m_b^3 \, f_B^2 \over 8
\pi} \left [ ({ C}_+^2-{ C}_-^2) \, { B} + {1 \over 3}\, ({ C}_+^2+{ C}_-^2) \,  { \tilde B} \right] 
\approx - 0.030 \, \left ( {f_B \over 220 \, MeV } \right )^2 \, ps^{-1},
\label{dgb}
\ee
where $f_B$ is the $B$ meson annihilation constant, the coefficients $C_+$ and $C_-$ are the standard non-leptonic weak interaction renormalization coefficients:
\be
C_-=C_+^{-2}=(\alpha_s(m_b)/\alpha_s(m_W))^{12/23}
\ee
and finally $B$ and $\tilde B$ are the `bag' constants defined for a generic pseudoscalar heavy meson $P$ as
\bea
&& \langle P_{Q \overline q} | (\overline Q \gamma_\mu (1-\gamma_5) q) \, (\overline q \gamma_\mu (1-\gamma_5) Q) | P_{Q \overline q} \rangle = {1 \over 2} \, f_P^2 \, M_P \, { B} \nonumber \\
&&\langle P_{Q \overline q} | (\overline Q \gamma_\mu (1-\gamma_5) Q) \, (\overline q \gamma_\mu (1-\gamma_5) q) | P_{Q \overline q} \rangle = {1 \over 6} \, f_P^2 \, M_P \, { \tilde B}~.
\label{bagb}
\eea
In the so-called factorization approximation (where the four-quark operators are replaced by the products of two-quark with the vacuum insertion), both bag constants are equal to one: $B= {\tilde B}=1$, in which approximation the numerical value of the decay rate difference in Eq.(\ref{dgb}) is estimated.

Experimentally measured differences of the total decay rates are: $\Gamma(B_s)-\Gamma(B_d^0)=0.027 \pm 0.013~ps^{-1}$ and $\Gamma(B^\pm) - \Gamma(B^0) = -0.043 \pm 0.006 ~ps^{-1}$, both in a reasonable agreement with the theoretical expectation, given the current errors. 

It can be also mentioned that the relative difference of the decay rates in the $B$ hadrons is quite small, while similar differences for the charmed mesons are quite conspicuous. In particular by scaling the effect to from the $B$ to $D$ mesons, one estimates that the difference in the lifetimes between the charged and the neutral $D$ mesons should be of order one:
\be
{\Delta \Gamma(D) \over \Gamma(D)}  \approx {m_b^2 \over m_c^2} \, {f_D^2 \over f_B^2} \, {\Delta \Gamma(B) \over \Gamma(B)} \sim O(1)~,
\ee
which is very well known to be the case indeed.

\subsection{$|V_{ub}|$ and nonfactorizable terms in inclusive decay rates}
The determination of the small mixing parameter $|V_{ub}|$ is complcated by the overwhelming background in the semileptonic decays arising from the dominant $b \to c$ transition. In order to separate the charmless semileptonic decays $B \to X_u \ell \nu$ one has to impose kinematical cuts that eliminate the contribution of the heavier charmed final states $X_c$. Such cuts however also leave only a minor fraction of the charmless decays as well. If, for example, the kinematical constrain is imposed on the value of $q^2$ of the lepton pair\cite{bauer00}, $q^2 > (M_B-M_D)^2$, one is left with only about 20\% of the total rate of the decays $B \to X_u \ell \nu$. The theoretical calculation of such fractional decay rate into a restricted part of the phase space however suffers from uncertaintees which substantially limit the accuracy of the extraction of the mixing parameter $|V_{ub}|$ from the data\cite{mv01}. In particular, within the OPE based approach the nonperturbative contribution of the third term in the effective Lagrangian (\ref{opem}) is formally concentrated in the `useful' part of the phase space. (One can also formulate this behavior by recognizing that the effective mass parameter for the expansion of the decay rate restricted by $q^2 > q_0^2$ is in fact $\mu = (m_b^2-q_0^2)/(2 m_b)$ which becomes less than the mass of the charmed quark if $q_0^2 = (M_B-M_D)^2$.)

The nonperturbative correction to the semileptonic decay due to the $b \to u$ current can be written as
\be
\delta^{(3)} \Gamma(B \to X_u \, \ell \, \nu) = 
{G_F^2 \, |V_{ub}|^2\, f_B^2 \, m_b^2 \, m_B \over 12 \, \pi} \, (B_2-B) \approx 0.3 (B_2-B) \, \Gamma(B \to X_u \, \ell \, \nu)~
\label{dbu}
\ee
where $B_2$ is still another bag constant defined as
\be
\langle B |{\bar b} (1-\gamma_5) u) ({\bar u} (1+\gamma_5) b) | B  \rangle = {f_B^2 \, m_B \over 2} \, B_2
\ee
and which also is equal to one in the heavy quark factorization limit. The actual values of the bag constants are currently not known. In the limit of large number of colors $N$ the deviations from the factorization limit are expected to be suppressed by $1/N^2$, which puts the expected difference between the bag constants at $O(0.1)$ in the real world. Clearly, an $O(3\%)$ uncertainty thus expected on the basis of Eq.(\ref{dbu}) is noticeable in comparison with the 20\% usable rate in the determination of $|V_{ub}|$. Thus an improvement in the precision of this parameter requires understanding the nonfactorizable terms.

To some extent the nonfactorizable contributions to the decay rates can be probed by the experimental data. One quantity explained by these contributions is the lifetime difference between the $D_s$ and $D^0$ mesons. The observed approximately 25\% diffrence agrees well with the understanding that the nonfactorizable terms should be of the order of 0.1 as compared to the factorizable ones (e.g. the difference in the lifetimes between the $D^\pm$ and $D^0$). A more direct access to the combination of the bag constants involved in Eq.(\ref{dbu}) is provided by the difference in the total {\it semileptonic} decay rates of the $D_s$ and $D^0$ mesons. Namely one can find that
\be
\Gamma_{sl}(D^0) \approx \Gamma_{sl}(D_s)=0.08 \, ps^{-1} \, \left ({m_c \over 1.4 \, GeV} \right )^2 \, \left ( {f_D \over 0.2 \, GeV} \right )^2 \, \left ( { \delta B_s \over 0.1} \right )~,
\label{dslds}
\ee
where $\delta B_s$ is the SU(3) nonsinglet part of the difference $(B_2-B)$ similar to the one in Eq.(\ref{dbu}). Clearly, if $\delta B_s \sim 0.1$ the relative effect of this difference should be quite conspicuous given that the total semileptonic decay rate of $D^0$ is $\Gamma_{sl}(D^0) \approx 0.165~ps^{-1}$. For this reason a measurement of the difference in inclusive semileptonic decay rates between the $D_s$ and $D^0$ is of a great interest and may significantly reduce the theoretical uncertainty in determination of the mixing parameter $|V_{ub}|$, which currently is understood~\cite{hfag} to be in the range $|V_{ub}|=(3 \div 5) \times 10^{-3}$.

\subsection{Some general considerations.}
The elements of the CKM matrix are currently known to a good extent. Of those that we not discussed here the parameter $|V_{ts}|$ is essentially fully fixed by the unitariry of the mixing matrix: $|V_{ts}| = |V_{cb}|$,
and this value is in a perfect agreement with the measured frequency of the $B_s - {\bar B}_s$ oscillations, $\Delta m_s = 17.8~ps^{-1}$, while the parameter $V_{td}$ is deduced from the $B_d - {\bar B}_d$ oscillation frequency, $\Delta m_d = 0.51~ps^{-1}$: $|V_{td}| \approx 8 \times 10^{-3}$, so that $|V_{td}|/|V_{ts}| \approx 0.20$. Combined with the measurement of $\sin 2 \beta$ from the CP-odd time asymmetry in the $B$ decays this in fact completes the test of the unitarity triangle and the measurement of its elements. 

It is often claimed that an improvement in the measurements of the CKM mixing parameters may probe a `New Physics'. Although this is a logical possibility, it can be noted that the accuracy of the determination of the weak-interaction parameters as well other quantities related to quarks (e.g. masses) still remains limited by the theoretical understanding of the `conversion' between the quark parameters and the properties of the observed hadrons. An illustrative example of this situation is provided by the original Cabibbo angle $\theta_c$, i.e. the CKM parameter $|V_{us}|$. The `gold plated' mode for measuring it is the $K_{e3}$ decay
$K \to \pi e \nu$. The form factor $f_+(0)$ for this decay is fixed in the limit of the flavor SU(3) symmetry and is protected from the first order corrections due to SU(3) breaking by the Ademollo-Gatto theorem. The Kaon decay data are abundant and readily available, and the history of the measurements spans more than four decades. However I believe that the uncertainty in the understanding of the second order SU(3) breaking effects places a substantial intrinsic limit on the accuracy of the knowledge of $\sin \theta_c$. The `natural' scale for the second order SU(3) violating effects is O(10\%). The Tables \cite{pdg} list the value of $|V_{us}|$ with a better than one percent error, based on the careful and detailed analysis of the value of $f_+(0)$ in Ref.\,\cite{leutwyler84}. The latter analysis however is based on the chiral perturbation theory and also uses a constituent quark model (on the light cone), which both have their limitations, especially if one aims at such a high accuracy. In particular some recent numerical (lattice) calculations do not confirm the results of the chiral analysis in \cite{leutwyler84}. (For a discussion see e.g. \cite{pdg}.) Thus the current accuracy of the knowledge of the Cabibbo angle is not entirely clear. On the other hand it is also not clear what dramatically new conclusions could be drawn had we known $\sin \theta$ with a much better precision. At least the present author is not aware of any credible theoretical prediction of this parameter, which would be thus tested by the data. Neither it looks very likely that improving the precision in the Cabibbo angle and other mixing parameters would lead us directly to a `New Physics'. After all --- when was the last time a `New Physics' has been found from the third (or even the second) significant digit in an otherwise known quantity, rather than from observing something qualitatively new?

\section{New hadronic dynamics}
Although extensive studies at the B factories have not brought any indications of a `New Physics' they did bring something unexpected and extremely interesting. Namely, it appears that a whole realm of unusual states containing have quarks has been uncovered that may significantly extend the understanding of hadronic dynamics. Until quite recently the reliably known hadrons have all fit the standard quark model template: either quark-antiquark mesons or three quark baryons. Anything not fitting this scheme, generically referred to as `exotics', was not seen with a convincing reliability, inspite of many searches. The situation has changed since the discovery\cite{bellex} of 
a narrow ($\Gamma < 2.3\,$MeV) resonance $X(3872)$ produced in the decays $B \to K \,
X$ and decaying as $X(3872) \to \pi^+ \pi^- \, J/\psi$ and with a mass in an extreme proximity to the $D^0 {\bar D}^{*0}$ threshold: $M_X-M(D^0 {\bar D}^{*0})=-0.6 \pm 0.6\,$MeV. By now it is pretty much clear that the 
$X(3872)$ is a near-threshold singularity in the $D^0 {\bar D}^{*0}$ S-wave channel with even C parity --- either a shallow bound state\,\cite{clopag,mvx,ess,nat}, or (more likely) a so-called virtual state\,\cite{bugg,hanhart,mvxv}. Such `molecular' states made of pairs of heavy-light hadrons were expected since long ago\,\cite{vo76,drgg}. Further studies at the B factories have uncovered new resonances with open and hidden charm, that exhibit exotic properties, although their internal dynamics is likely to be different from the `molecular' structure of $X(3872)$. In the remaining part of this lecture I discuss some properties of the $X(3872)$ and also of the so-called $Y$ and $Z$ resonances decaying into specific states of charmonium and pions, which may be states of an entirely new type, namely, of charmonium bound inside an excited light hadronic matter, `hadro-charmonium'\,\cite{mvch}.

\subsection{$X(3872)$}
The state $X(3872)$ observed in $B$ decays\,\cite{bellex,babarx} and in $p \bar p$ collisions at the Tevatron\,\cite{cdfx,d0x}. The co-existence of the decay $X \to \pi^+  \pi^- J/\psi$ (the discovery mode) on one hand and the decays $X \to \pi^+  \pi^- \pi^0 J/\psi$ and $X \to \gamma J/\psi$ on the other implies that the isospin is badly broken in the $X(3872)$. The C parity of the resonance has to be positive, and the bulk of data indicates that the quantum numbers of this state are $J^{PC}=1^{++}$. These properties can be explained if $X$ is a C-even S-wave state of a charmed meson pair $D^0 {\bar D}^{*0}+ {\bar D}^0 D^{*0}$. The threshold for a similar pair of charged mesons, $D^+ D^{*-}+ D^- D^{*+}$ is heavier by $\delta \approx 8\,$MeV which is a large gap in the scale of the splitting of $X$ from the $D^0 {\bar D}^{*0}$ threshold. The presence in the wave function of the pair of charged mesons is thus heavily suppressed, which gives rise to $X$ being a mixture of isospin 0 and isospin 1 isotopic states. If $X$ is dominantly a shallow bound state of the neutral mesons, those mesons should move inside the resonance at distances beyond the range of strong interaction, similarly to the proton and the neutron bound in a deutron. In this picture one should expect that the know decays of the $D^{*0}$, or the ${\bar D}^{*0}$ meson from the `peripheral'
component of the $X(3872)$, $D^{*0} \to \gamma D^0$, $D^{*0} \to \pi^0 D^0$, give rise to the decays of the $X$ resonance $X \to \gamma D^0 {\bar D}^0$ and $X \to \pi^0 D^0 {\bar D}^0$ with a calculable rate and an interference pattern determined by the binding energy\cite{mvx,mvxd}. An experimental search for these decays produced an unexpected result: the peak in the $\gamma D^0 {\bar D}^0$ and $\pi^0 D^0 {\bar D}^0$ invariant mass spectra in the decays $B \to X K$ turned out to be at or slightly above the threshold for $D^0 {\bar D}^{*0}$ and was consistent with decays of unbound $D^{*0}$ mesons\cite{belledd,babardd}. This behavior suggests\,\cite{hanhart} that $X(3872)$ is in fact a so-called virtual S-wave state in the C-even $D^0 {\bar D}^{*0}+ {\bar D}^0 D^{*0}$ channel, i.e. it corresponds to a large negative scattering length for the meson scattering near the threshold. In this case the scattering of the mesons near the threshold can be analyzed in both isotopic channels with $I=0$ and $I=1$, and the interplay between the kinematics involving two thresholds for the $D^0 {\bar D}^{*0}$ and $D^+ D^{*-}$ states and the isospin symmetry of the underlying strong interaction produces a distinctive pattern of the dependence of the isospin structure of the scattering states on the energy $E$ relative to the threshold\,\cite{mvxv}. This pattern can be tested by observing a different behavior of the threshold peaks in the $X \to \pi^+  \pi^- J/\psi$ ($I=1$) and the 
$X \to \pi^+  \pi^- \pi^0 J/\psi$ channels in the processes $B \to X K$ as illustrated in Figure~2.

\begin{figure}[tb]
  \begin{center}
\begin{minipage}[t]{18 cm}
\epsfxsize=9cm
    \epsfxsize=9cm
     \epsfbox{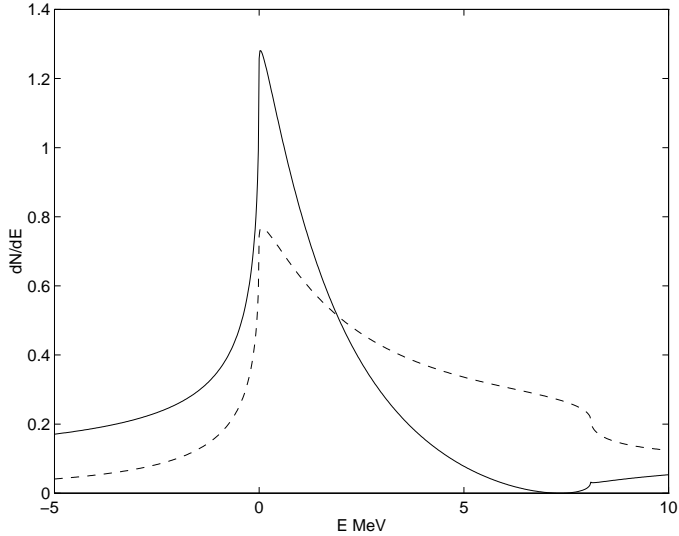}
\end{minipage}
\begin{minipage}[t]{16.5 cm} 
     \caption{The expected shape (in arbitrary units) of the virtual state peak
in the yield of $\pi^+ \pi^- J/\psi$ (solid) and $\pi^+ \pi^- \pi^0 J/\psi$
(dashed) channels. The energy $E$ (invariant mass) is measured from the $D^0 {\bar D}^{*0}$ threshold. \label{xspectr}}
\end{minipage}
\end{center}
\end{figure}

\subsection{$Y, Z$ --- hadro-charmonium?}
The charmonium resonances with mass below the open charm threshold ($J/\psi, \psi', \chi_{cJ}, \ldots $) are narrow due to suppression of the annihilation of a heavy quark pair into light hadrons. This suppression, known as the Okubo-Zweig-Iizuka (OZI) rule also finds its quantitative explanation in QCD~\cite{ap}. Conversely, for higher charmonium resonances, i.e. those above the open charm threshold, the decay into pairs of charmed mesons should not be suppressed, and those resonances should be broad. The known charmonium states indeed generally follow this rule, e.g. $\psi(3770)$ is only $30 \div 40$\,MeV above the $D {\bar D}$ threshold, and its width is about 25\,MeV and is essentially all due to the decay into pairs of charmed mesons. However notable exceptions from such pattern have been recently found. The resonance $Y(4260)$ with quantum numbers $1^{--}$ has a large width of about 80\,MeV, but non of its decays into charmed mesons have ever been seen. Rather the observed decay modes are $Y \to \pi \pi J/\psi$ and $Y \to K {\bar K} J/\psi$~\cite{babar4260,belle4260,cleo4260,cleo4260a}. Subsequently more $1^{--}$ peaks with similar behavior were observed in the range 4.32 - 4.36\,MeV, $Y(4.32-4.36)$~\cite{babar432,belle432} and near 4.66\,MeV, $Y(4.66)$~\cite{belle432}, corresponding to the decay $Y \to \pi \pi \psi'$ (but not $J/\psi$). Finally, a manifest isotopic non-singlet  $Z(4430)$ has been reported\cite{bellez} (but not yet independently confirmed~\cite{babarz}) in $B$ decays, corresponding to a resonance in the $\pi^\pm \psi'$ channel, followed by an indication~\cite{bellez1} of similar peaks $Z_1(4.05)$ and $Z_2(4.25)$ decaying into $\pi^\pm \chi_{c1}$.

The apparent existence of resonances that decay into a specific low-mass charmonium state and light hadrons invites an explanation~\cite{mvch} that these resonances in a sense already contain inside them that charmonium specific state. In other words, these high-mass resonances contain charmonium embeded in an excited light hadronic matter, so that in their observed decays the light degrees of freedom are `shaken off'. Generally the idea of a binding of charmonium in a light matter is not novel --- bound states of $J/\psi$ inside nuclei has been discussed for some time~\cite{bst,vs}. The novel feature of the structure of the possible new resonances, which we call hadro-charmonium, is that the ligh-matter `host' for the charmonium is an excited (and unstable) state rather than a stable nucleus. The existence of such states can be understood in the limit of heavy quarkonium and of a highly excited light-matter resonance by considering a van der Waals type interaction between the quarkonium and the light matter~\cite{mvch,dv}.

The interaction of a compact charmonium state (call it generically $\psi$) with long wave length gluonic field inside light hadronic matter can be described in terms of the multipole expansion in QCD\cite{gottfried,mv79,peskin} with the leading term being the E1 interaction with the chromo-electric field ${\vec E}^a$. The effective Hamiltonian arising in the second order in this interaction can then be written as
\be
H_{\rm eff}= -{1 \over 2} \, \alpha^{(\psi)} \, E_i^a E_i^a~,
\label{hdiag}
\ee
where $\alpha^{(\psi)}$ is the (diagonal) chromo-electric polarizability of the state $\psi$ having the dimension of volume, and which can be expressed in terms of the Green's function ${\cal G}$ of the heavy quark pair in a color octet state:
\be
\alpha^{(\psi)} = {1 \over 16} \langle \psi | \xi^a r_i {\cal G} r_i \xi^a | \psi \rangle~,
\label{alpsi}
\ee
with ${\vec r}$ being the relative position of the quark and the antiquark and $\xi^a$ the difference of the color generators acting on them: $\xi^a = t_c^a- t_{\bar c}^a$ (a detailed discussion can be found e.g. in the review \cite{mvch}). Finally, the QCD coupling $g$ is included in the normalization of the field strength in Eq.(\ref{hdiag}), so that e.g. the gluon Lagrangian in this normalization takes the form $-(1/4 g^2) F^2$.

The diagonal chromo-electric polarizability is not yet known for either of the charmonium states, although it can be directly measured for the $\jp$ resonance\cite{mvpol}. What is known is the off-diagonal chromo-polarizability $\alpha^{(\psi \psi')}$ describing the strength of the amplitude of the transition $\pp \to \pi^+ \pi^- \jp$: $|\alpha^{(\psi \psi')}| \approx 2 \,$ GeV$^{-3}$~\cite{mvch,vs}. On general grounds, the diagonal chromo-polarizability for each of the $\jp$, $\pp$ and $\chi_c$ states should be real and positive, and for the former two the Schwartz inequality 
\be
\alpha^{(J/\psi)} \, \alpha^{(\psi')} \ge |\alpha^{(\psi \psi')}|^2
\label{schw}
\ee
should hold. It can be expected however that each of the discussed diagonal parameters should exceed the off-diagonal one, and also that the chromo-polarizability for the $\pp$ and $\chi_c$ states is larger than that of the $\jp$ due to their larger spatial size.

The strength of the van der Waals type interaction between a charmonium state and a hadron (generically denoted here as $X$) can be evaluated using the effective Hamiltonian (\ref{hdiag}) and the well known expression for the conformal anomaly in QCD in the chiral limit:
\be
\theta_\mu^\mu = -{9 \over 32 \pi^2} \, F_{\mu \nu}^a F^{a \,\mu \nu}= {9 \over 16 \pi^2} \, \left ( E_i^a E_i^a - B_i^a B_i^a \right )~,
\label{anom}
\ee
where ${\vec B}^a$ is the chromo-magnetic field, and the (normalization) expression for the static average of the trace of the stress tensor $\theta_\mu^\mu$ over any state $X$ in terms of its mass $M_X$:
\be
\langle X | \theta_\mu^\mu ({\vec q}=0) | X \rangle = M_X~,
\label{mx}
\ee
which is written here assuming nonrelativistic normalization for the state $X$:
$\langle X | X \rangle=1$. Averaging the effective Hamiltonian (\ref{hdiag}) over a hadron $X$ made out of light quarks and/or gluons, one thus finds 
\be
\langle X | H_{\rm eff} | X \rangle \le - {8 \pi^2 \over 9} \, \alpha^{(\psi)} \, \, M_X~,
\label{ieff}
\ee
where the inequality arises from the fact that the average value of the full square operator $B_i^a B_i^a$ over a physical hadron $X$ has to be non-negative.
The relation (\ref{ieff}) shows an integral strength of the interaction. Namely, if the force between the charmonium ($\psi$) and the light hadronic matter inside the hadron $X$ is described by a potential $V({\vec x})$, such that $V$ goes to zero at large $|{\vec x}|$, this relation gives the bound for the integral
\be
\int V({\vec x}) \, d^3x \le - {8 \pi^2 \over 9} \, \alpha^{(\psi)} \, M_X~.
\label{iint}
\ee

The value of the integral in Eq.(\ref{iint}), although undoubtedly corresponding to an attraction, does not by itself automatically imply existence of a bound state, since it does not take into account the kinetic energy, which in a nonrelativistic treatment of the system is $p^2/2{\bar M} \sim 1/(R^2 \bar M)$, with $R$ being a characteristic size of the hadron $X$ and $\bar M = M_X M_\psi/(M_X+M_\psi)$ the reduced mass in the system. The spatial integral of the kinetic energy is then parametrically of order $R/{\bar M}$, and, given the relation in Eq.(\ref{iint}), the condition for existence of a bound state can be written as
\be
\alpha^{(\psi)} \, {M_X \, \bar M \over R} \ge C~,
\label{cond}
\ee
where $C$ is a numerical constant (parametrically of order one) which depends on a model for the distribution of the interaction over the interior of the hadron $X$ and thus on a more precise definition of the ``characteristic size" $R$.
Considering the condition (\ref{cond}) for excited light-matter resonances, one can readily see that the appearance of a hadro-charmonium state, possibly at a higher excitation, depends on the behavior of the combination $M_X \, \bar M / R$ with the excitation number of the resonance $X$. In particular, if the characteristic size $R$ grows slower than the mass $M_X$ a binding of charmonium necessarily occurs in a sufficiently highly excited resonance. Such behavior takes place, e.g. in the once popular bag model\cite{bagm}, where $R \propto M^{1/3}$. However in models which reproduce the approximately linear behavior of the Regge trajectories for the resonances, such as string model including its recently discussed\cite{kkss} implementation in terms of AdS/QCD correspondence with `linear' confinement, one effectively finds $R \propto M$, and a more accurate estimate~\cite{dgv} proves that the binding of heavy quarkonium necessarily occurs at sufficiently high excitation of the light hadronic matter. Also one can argue~\cite{dgv} that in the limit of large mass $m_Q$ of the heavy quark the decay of such states into open heavy flavor channels, requiring a reconnection of the heavy-heavy and light-light binding, is suppressed as $\exp(-{\rm const.} \, \sqrt{\Lambda_{QCD}/m_Q} )$.

Obviously, the existing idealized theoretical picture can provide only qualitative guidance for the properties of the real-life hadro-quarkonium states. The most straightforward of these qualitative predictions are the following:\\
\begin{itemize}
\item{In addition to mesonic hadro-charmonium resonances there should exist bound states of baryo-charmonium, i.e. bound states of $J/\psi$, $\psi'$, $\chi_c$ with baryonic excitations at low baryon number. Such states should decay into e.g. a nucleon + $J/\psi$, possibly with additional pions.}
\item{There should exist resonances decaying into pion(s) plus other low-mass states of charmonium, $\eta_c$, $\chi_c$. This prediction~\cite{mvch,dv} is in a qualitative agreement with the apparent observation of $Z_{1,2}$~\cite{bellez1}.}
\item{In addition to the observed decays into final states with the `preferred' specific charmonium state, e.g. $Y(4260) \to \pi \pi J/\psi$, the hadro-charmonium resonances should also decay, at a smaller rate, to non-preferred states, e.g. $Y(4260) \to \pi \pi \psi'$ due to a `deformation' of charmonium in the bound state.}
\item{A similar to hadro-charmonium structure of resonances should exist for the bottomonium in the range around 11 - 11.5\,GeV, most likely with the excited low-mass bottomonium states (due to a larger chromo-polarizability of the excited states). A possible hadro-bottomonium structure is in fact hinted by the data\,\cite{belleups} on the $\pi \pi \Upsilon(nS)$ yield near 10.9\,GeV.}
\end{itemize}    

\section*{Acknowledgments}
This work is supported in part by the DOE grant DE-FG02-94ER40823.\\[0.1in]

\end{document}